      [1999/12/01 v1.4c Il Nuovo Cimento]
\documentclass{cimento}
\usepackage{graphicx}
\usepackage{subfigure}

\title{Final measurement of \Bs\ mixing phase in the full CDF Run II data set}
\author{
  S.~Leo\from{ins:x} on behalf of the CDF collaboration.
}
\instlist{
  \inst{ins:x} University and INFN of Pisa - Pisa, Italy
}
\PACSes{
13.25.Hw   % Hadronic decays of bottom mesons
11.30.Er   % Charge conjugation, parity, time reversal, and other discrete symmetries 
14.40.Nd   % Bottom mesons
14.65.Fy   % from old PRL
12.38.Qk   % Experimental tests 
}

\def\B0    {\ensuremath{B^0}}

\def\Bs    {\ensuremath{B^0_s}}
\def\bBs  {\ensuremath{\bar{B}^0_s}}

\def\Jpsi{\ensuremath{J\!/\!\psi}}
\newcommand{\myto}{\kern -0.3em\to\kern -0.2em}
\def\BsJpsiPhi{\ensuremath{\Bs \myto \Jpsi \phi}}
\def\betas     {\ensuremath{\beta_s}}
\def\DGs       {\ensuremath{\Delta\Gamma_s}}

\def\taus   {\ensuremath{\tau_{s}}}

\newcommand{\TeV}{\ensuremath{\matqq	hrm{Te\kern -0.1em V}}}
\newcommand{\TeVc}{\ensuremath{\mathrm{Te\kern -0.1em V\!/}c}}
\newcommand{\TeVcc}{\ensuremath{\mathrm{Te\kern -0.1em V\!/}c^2}}
\newcommand{\GeV}{\ensuremath{\mathrm{Ge\kern -0.1em V}}}
\newcommand{\GeVc}{\ensuremath{\mathrm{Ge\kern -0.1em V\!/}c}}
\newcommand{\GeVcc}{\ensuremath{\mathrm{Ge\kern -0.1em V\!/}c^2}}
\newcommand{\MeV}{\ensuremath{\mathrm{Me\kern -0.1em V}}}
\newcommand{\MeVc}{\ensuremath{\mathrm{Me\kern -0.1em V\!/}c}}
\newcommand{\MeVcc}{\ensuremath{\mathrm{Me\kern -0.1em V\!/}c^2}}

\newcommand{\ppbar}{\ensuremath{p\overline{p}}}

\newcommand{\KpKm}{\ensuremath{K^+K^-}}

\newcommand{\CP}{\ensuremath{C\!P}}

\renewcommand{\epsilon}{\varepsilon}
\renewcommand{\theta}{\vartheta}

\begin{document}

\maketitle

\begin{abstract}

We report the final CDF measurement of the \Bs\ mixing phase, mean lifetime, and decay-width difference through the fit of the time 
evolution of flavor-tagged \BsJpsiPhi\ decays.
The measurement is based on the full data set of 1.96 TeV \ppbar\ collisions collected between February 2002 and September 2011 by the CDF experiment. 
The results are consistent with the standard model and other experimental determinations and are amongst the most precise to date.

\end{abstract}

\section{Introduction}

Flavor physics of quarks is considered one of the most promising sectors for indirect signs of new particles or 
interactions beyond the standard model (SM). The \Bs\ dinamics in particular offers rich opportunities since 
its experimental exploration has not reached in extension and precision stringent constraints on the presence of NP in contrast to what happened in leading (and some subleading) processes 
involving charged and neutral kaons and bottom mesons.
The \Bs\ oscillations are explained in terms of second-order weak processes involving the the CKM matrix element $V_{ts}$:
 a broad class of generic extensions of the SM is expected to affect the mixing amplitude, modifying the mixing ``intensity''--that is 
the oscillation frequency--and the phase, \betas. A non-SM enhancement of \betas\  would also decrease the size of 
the decay-width difference between the light and heavy mass eigenstates of the \Bs\ meson, \DGs~\cite{Faller:2008gt}. 
While the oscillation intensity has been measured precisely~\cite{Abulencia:2006ze}, only loose constraints on the phase were available until recently. 
The most effective determination of \betas\ and \DGs\ is achieved through the analysis of the time evolution of flavor--tagged \BsJpsiPhi\ decays.
% Since the decay is dominated by a single real amplitude, the phase difference equals the mixing phase, \betas, 
% to a good approximation~\cite{Faller:2008gt}.
% Studies of the phase difference between the \Bs-\bBs\ mixing amplitude and the amplitude of \Bs\ and \bBs\ decays into
% common final states are considered very powerfull in probing the presence of physics beyond SM contributions in $V_{ts}$.
% The values of the mixing phase and width difference are loosely constrained, and currently
% the subject of intense experimental activity.
% The most effective determination of \betas\ and \DGs\ can be achieved through the analysis of time evolution 
% of \BsJpsiPhi\ decays.
% The first determinations of \betas, by the CDF and D0 experiments, suggested a mild deviation from the SM expectation~\cite{sin2betas-early}. 
% However, updated measurements in \BsJpsiPhi\ decays~\cite{CDF:2011af,Abazov:2011ry,LHCb:2011aa} showed increased 
% consistency with the SM, calling for additional experimental information to clarify the picture.\par

The first such analysis was performed by the CDF experiment in 2008~\cite{sin2betas-early}. 
Immediately after D0 joined. In 2010 the combination of CDF and D0 results suggested a mild deviation from the SM expectation.
However, updated measurements ~\cite{CDF:2011af,Abazov:2011ry,LHCb:2011aa} showed increased 
consistency with the SM, calling for additional experimental information to clarify the picture.\par
Here we report the new CDF update using the final data set of 10 fb$^{-1}$.%~\cite{phis_cdf_new}. 

%%%%% (FINE PARTE DA RISCRIVERE DA SCRATCH)

\section{Analysis}
\begin{figure}[hbtp]
\begin{center}
\includegraphics[width=0.45\textwidth,height=0.45\textwidth]{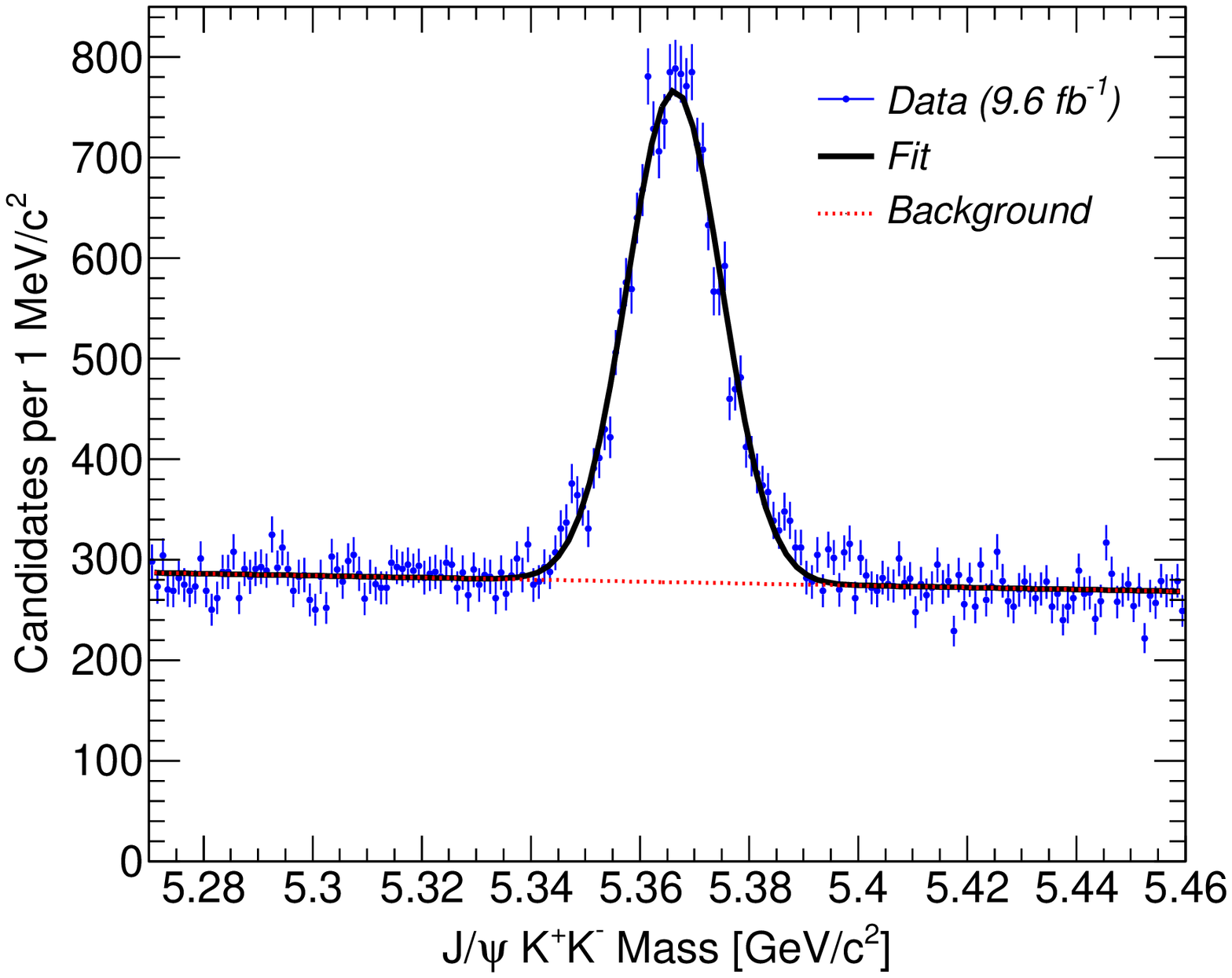}
\includegraphics[width=0.45\textwidth,height=0.45\textwidth]{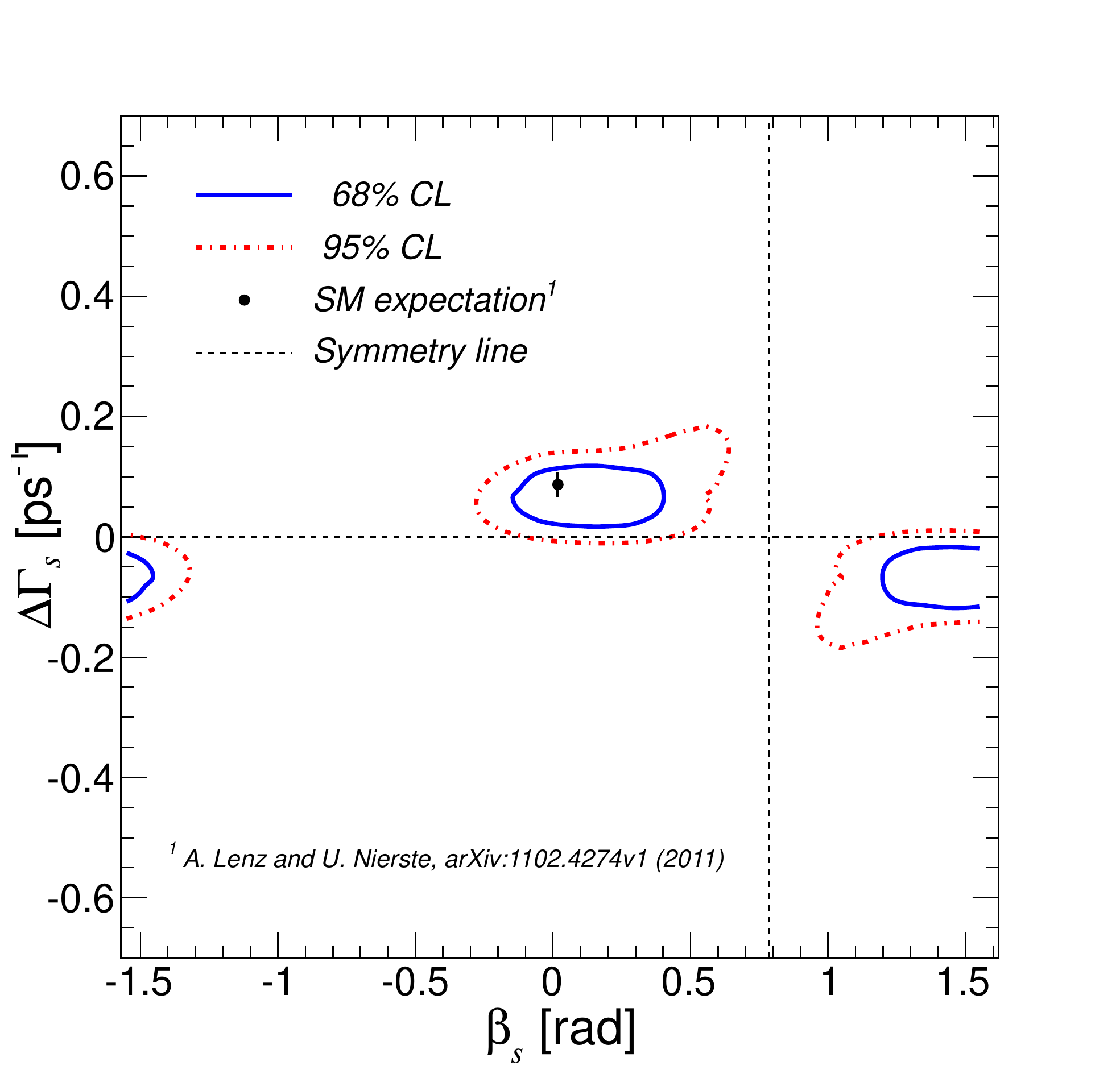}
\end{center}
\caption{\textbf{Left:} Distribution of $\Jpsi K^+K^-$ mass with fit projection overlaid. \textbf{Right:} Confidence regions at the 68\% 
and 95\% CL in the (\betas, \DGs) plane.}
\label{fig:mass}
\label{fig:contour}
\end{figure}

%%%%%%%%%%%%%%%%%%%%%%%%%%%%%%%%%%%%%%%%%%%%%%%%%%%%%%%%%%%%%%%%%%%%%%%%%%%%%
%Detector and data set
%%%%%%%%%%%%%%%%%%%%%%%%%%%%%%%%%%%%%%%%%%%%%%%%%%%%%%%%%%%%%%%%%%%%%%%%%%%%%
% The CDF II detector~\cite{CDF:detector} is a magnetic spectrometer 
% surrounded by electromagnetic and hadronic calorimeters and muon detectors that has cylindrical geometry with
% forward-backward symmetry. Charged particle trajectories (tracks) are reconstructed with resolution on the momentum component transverse to the beam ($p_T$) of $\sigma_{p_{T}}/p_{T} \approx 0.07\%/p_T$ ($p_T$ in GeV/$c$), corresponding to a mass resolution for our signals of about 9 MeV/$c^2$. 
The decays, collected by a low--momentum dimuon trigger, are fully reconstructed using four tracks originating from a common displaced vertex, 
two matched to muon pairs consistent with a \Jpsi\ decay ($3.04< m_{\mu\mu}< 3.14$ \GeVcc), and two consistent with a $\phi \myto \KpKm$ decay ($1.009< m_{KK}< 1.028$ \GeVcc).
The dimuon mass constraint to the known \Jpsi\ mass combined with a transverse momentum resolution of $\sigma_{p_{T}}/p_{T} \approx 0.07\% p_T$ ($p_T$ in GeV/$c$)
yield a mass  resolution for our signals of about 9 MeV/$c^2$.
The $\Jpsi K^+K^-$ mass distribution (fig.~\ref{fig:mass}, left), shows a signal of approximately 11~000 decays, overlapping a similar amount of constant background dominated by the prompt combinatorial component, and smaller contributions from mis-reconstructed $B$ decays.

% \section{Analysis Strategy and tools}
The analysis relies on a joint fit to the time evolution of \Bs\ mesons that resolve the fast oscillations by exploiting 
the 90~fs time resolution of the CDF micro--vertex detector for these final states.
% Decays of states initially produced as a  $B^0_s$ or $\overline{B}^0_s$ meson are treated independently as well as decays that result in a \CP-odd or \CP-even combination of the $\Jpsi \phi$ angular momenta. 
Because the $B^0_s$ meson has spin zero and \Jpsi\ and $\phi$ have spin one,   
the \BsJpsiPhi\ decay involve three independent amplitudes, each corresponding to one possible angular momentum state 
of the $\Jpsi \phi$ system which is also a \CP-odd or \CP-even eigenstate.
To enhance the sensitivity to \betas, the time-evolution of the decay amplitudes is fitted independently by exploiting differences 
in the distribution of the kaons' and muons' decay angles.
% These are used in the fit through the transversity basis~\cite{Dighe:1998vk}, 
% which allows a convenient separation of the various terms in the likelihood.
% The time evolution of an initially produced $\Bs$~meson differs from that of a $\bBs$~meson. 
% Accounting for this difference in the fit further enhances the sensitivity to \betas. 
Sensitivity to \betas\ can be further enhanced accounting for the difference in the time evolution of initially produced \Bs\ and \bBs.
The flavor of the meson at the time of production is inferred by two independent classes of algorithms:
the opposite-side flavor tag (OST) and the same-side kaon tag (SSKT)~\cite{Abulencia:2006ze}.

The OST performances have been determined with 82~000 $B^\pm \myto \Jpsi(\myto \mu^+\mu^-) K^\pm$ decays fully reconstructed in the same 
sample as the signal. 
We found an efficiency of $\epsilon_{\rm{OST}} = (92.8\pm0.1)\%$, an observed averaged dilution, $D_{\rm{OST}} = 1 -2 w_{\rm{OST}}$, equal to $(12.3 \pm 0.4)\%$ and a resulting effective tagging power of $\epsilon_{\rm{OST}}D^2_{\rm{OST}} = (1.39 \pm 0.05)\%$. 
The SSKT algorithms tag a smaller fraction of candidates with better precision. 
Its performances have been previously determined~\cite{CDF:2011af} to be $\epsilon_{\rm{SSKT}} = (52.2 \pm 0.7)\%$, $D_{\rm{SSKT}}=(21.8\pm 0.3)\%$ 
and $\epsilon_{\rm{SSKT}}D^2_{\rm{SSKT}}=(3.2 \pm 1.4)\%$.\par
Since SSKT algorithm has been calibrated for early data only, we conservatively restrict its use to the events collected in that period. 
Simulation shows that this results in a modest degradation in \betas\ resolution.\par
The unbinned maximum likelihood joint fit uses 9 observables from each event to determine 32 parameters including $\beta_s$ and $\Delta\Gamma$,  other physics parameters (\Bs\ lifetime, decay amplitudes at $t=0$ and phases, etc),  and several other (``nuisance") parameters (experimental scale factors, etc.).
\section{Results}

If \betas\ is fixed to its SM value, the fit shows unbiased estimates and gaussian uncertainties for \DGs,
\taus. 
We found $\DGs =  0.068 \pm 0.026(stat) \pm0.007(syst)$ ps$^{-1}$, and mean \Bs\ lifetime,  $\tau_s = 1.528 \pm 0.019(stat) \pm0.009(syst)$ ps.
Systematic uncertainties include mismodeling of the signal mass model, lifetime resolution,  acceptance description, and angular distribution of the background;
a $\mathcal{O}(2\%)$ contamination by $\B0 \myto \Jpsi K^*(892)^0$ decays misreconstructed as \BsJpsiPhi\ decays; and the silicon detector misalignment.
% The uncertainty on the \DGs\ measurement is dominated by the mismodeling of the background decay time. The largest contribution to the uncertainty on \taus\ is the effect of  
% silicon detector misalignment. \par
These results are among the most precise measurements from a single experiment.

If \betas\ is free to float in the fit, tests in statistical trials show that the maximum likelihood estimate 
is biased for the parameters of interest, and the biases depend on the true values 
of the parameters. Hence, we determine confidence regions in the \betas\ and $(\betas,\DGs)$ spaces (fig.~\ref{fig:contour}, right), by using a profile-likelihood ratio statistic as a $\chi^2$ variable and considering all other likelihood variables as nuisance parameters.
Confidence regions are corrected for non-Gaussian likelihood and systematic uncertainties such as to ensure nominal coverage.
By treating \DGs\ as a nuisance parameter, we also obtain $\beta_s\in[-\pi/2,-1.51]\cup[-0.06,0.30]\cup [1.26,\pi/2]$ rad at the 68\% CL, and $\betas \in [-\pi/2, -1.36] \cup [-0.21,0.53] \cup [1.04,\pi/2]$ rad at the 95\% CL.
Included in the fit is also the CP-odd component originated by non resonant $K^+K^-$ pair or by the $f_0(980)$ decays. 
The resulting S-wave decay amplitude is found to be negligible.
All results are consistent with the SM expectation and with determinations of the same quantities from other experiments~\cite{Abazov:2011ry,LHCb:2011aa,Aaij:2012eq}.

% \begin{figure}
% \includegraphics{figures/2d_contour_yextended}     % includes figure foo.eps
% \caption{Confidence regions at the 68\% and 95\% CL in the (\betas,\DGs) plane.}
% \end{figure}

% \section{Summary} 
% In summary final CDF results on the \Bs\ mixing phase and
% decay width difference from the time-evolution of flavor-tagged $\BsJpsiPhi$ decays reconstructed in the full Tevatron Run II data set is reported.
% This analysis improves and supersedes the previous CDF measurement obtained in a subset of the present data \cite{CDF:2011af}.
% Considering \DGs\ as a nuisance parameter, and using the recent determination of the sign of \DGs~\cite{Aaij:2012eq}, $-0.06< \beta_s< 0.30$ rad at the 68\% CL has been found.
% Assuming a SM value for \betas, precise measurements of width difference,  $\DGs =  0.068 \pm 0.026(stat) \pm0.007(syst)$ ps$^{-1}$, and 
% mean \Bs\ lifetime,  $\tau_s = 1.528 \pm 0.019(stat) \pm0.009(syst)$ ps are also reported.
% All results are consistent with the SM expectation and with determinations of the same quantities from other experiments~\cite{Abazov:2011ry,LHCb:2011aa}.
% \appendix

% \acknowledgments


\begin{thebibliography}{0}

% \bibitem{hfag}
%   \BY{Asner~D \atque others}  
%   \NAME{arXiv:1010.1589} and online update at http://www.slac.stanford.edu/xorg/hfag.

\bibitem{Faller:2008gt}
  \BY{Faller~S., Fleischer~R. \atque Mannel~T.}
  \IN{Phys.\  Rev.\  D}{79}{2009}{014005}.

\bibitem{Abulencia:2006ze}
  \BY{Abulencia~A. \atque others}
  \NAME{CDF Collaboration},
  \IN{Phys.\ Rev.\ Lett.}{97}{2006}{242003}

\bibitem{sin2betas-early}
  \BY{Aaltonen~T. \atque others} 
  \NAME{CDF Collaboration}, 
  \IN{Phys.\  Rev.\  Lett.}{100}{2008}{161802}; 
  \BY{Abazov~V.M. \atque others}
  \NAME{D0 Collaboration},
  \IN{Phys.\  Rev.\  Lett.}{101}{2008}{241801}

% \bibitem{asl}
%   \BY{Abazov~V.M. \atque others}
%   \NAME{D0 Collaboration},
%   \IN{Phys.\  Rev.\  D}{82}{2010}{082001};
%   \SAME{84}{2011}{052007} 

\bibitem{CDF:2011af}
  \BY{Aaltonen~T. \atque others} 
  \NAME{CDF Collaboration},
  \IN{Phys.\ Rev.\  D}{85}{2012}{072002}.

\bibitem{Abazov:2011ry}
  \BY{Abazov~V.M. \atque others}
  \NAME{D0 Collaboration},
  \IN{Phys.\ Rev.\  D}{85}{2012}{032006},
  \SAME{82}{2010}{082001}, \SAME{84}{2011}{052007}


\bibitem{LHCb:2011aa}
  \BY{Aaij~R. \atque  others}
  \NAME{LHCb Collaboration},
  \IN{Phys.\ Rev.\  Lett.}{108}{2012}{101803}

% \bibitem{CDF:detector}
%   \BY{Blair~R. \atque others}
%   \NAME{CDF Collaboration}, 
%   \TITLE{The CDF II detector: Technical design report}, 
%   \NAME{FERMILAB-DESIGN-1996-01,1996}

% \bibitem{pdg} 
%   \BY{Nakamura,K. \atque others}
%   \NAME{Particle Data Group},
%   \IN{J.\ Phys.\ G}{37}{2010}{075021} and 2011 partial update for the 2012 edition.

% \bibitem{Dighe:1998vk}
%   \BY{Dighe~A.S., Dunietz~I. \atque and Fleischer~R.}      
%   \TITLE{Extracting CKM phases and $B_s - \bar{B}_s$ mixing parameters from angular distributions of nonleptonic $B$ decays},
%   \IN{Eur.\ Phys.\ J.}{C6}{1999}{647}


\bibitem{Aaij:2012eq}
  \BY{Aaij~R. \atque  others}
  \NAME{LHCb Collaboration},
  \TITLE{arXiv:1202.4717 (2012)}.

% \bibitem{phis_cdf_new} 
%   \BY{T. Aaltonen \atque others} 
%   \NAME{CDF Collaboration}, 
%    arXiv:xxxxxxxx.

\end{thebibliography}
\end{document}